\documentclass[twocolumn]{article} 
\pdfoutput=1
\usepackage[utf8]{inputenc}

 \usepackage{color, float}
 \usepackage{subfig}
 \usepackage{graphicx}
 \usepackage{amsmath,amssymb}
 \usepackage{tikz}
 \usepackage[labelfont=bf]{caption}
 
 \usepackage{multirow}

 \newcommand{\chv}{\ensuremath{\mathbf{v}}} 
 \newcommand{\bfx}{\ensuremath{\mathbf{x}}}

 \newcommand{\bfX}{\ensuremath{\mathbf{X}}}

\begin{document}
 
\title{Generalized Method of Moments for Estimating Parameters of Stochastic Reaction Networks}

  \author{Alexander L\"uck, Verena Wolf\\
  \{alexander.lueck, verena.wolf\}@uni-saarland.de\\
  Department of Computer Science, Saarland University, 66123 Saarbr\"ucken, Germany}
\twocolumn[
  \begin{@twocolumnfalse}
     \date{~}
    \maketitle
    \begin{abstract}
Discrete-state stochastic models  
have become a well-established approach to describe biochemical reaction networks
 that are influenced by the inherent randomness of cellular events.
In the last years  several  methods for accurately
approximating the statistical moments of such models
have become very popular since they allow an efficient analysis of complex networks.
 We propose a generalized method of moments approach
  for inferring the parameters of reaction networks 
based on a sophisticated matching of the statistical moments
of the corresponding stochastic model 
 and the sample moments of   population snapshot data. 
The proposed parameter estimation method exploits recently developed 
moment-based approximations and provides
  estimators with desirable statistical properties when a large number of samples is available.
 We  demonstrate the usefulness and
efficiency of the inference method on two case studies.
The generalized method of moments provides accurate and fast estimations of 
unknown parameters of reaction networks. The accuracy   
increases when also moments of order higher than two are considered.
In addition, the variance of the estimator decreases,
when more samples are given 
 or  when higher order moments are included.\vspace*{5ex}
    \end{abstract}
  \end{@twocolumnfalse}]

\section{Background}
A widely-used  approach in systems biology research is to design
quantitative models of biological processes and refine them based on both computer simulations
and wet-lab experiments. 
While a large amount of sophisticated parameter inference methods 
have been proposed for deterministic models, 
 only few approaches allow the efficient  calibration of parameters for 
 large discrete-state  stochastic models  that describe   stochastic interactions between
  molecules within a single cell.
Since research progress in experimental measurement
  techniques that deliver  single-cell and single-molecule data
  has   advanced, the ability to 
  calibrate   such models is of key importance.
For instance, the widely-used   flow cytometric analysis
delivers  data from thousands of cells which yields  sample means and
sample variances of molecular populations.   
Here, we focus on the most common scenario: a discrete stochastic model
of a cellular reaction network with unknown reaction rate constants and population
snapshot data such as sample moments of a large number of
observed samples.
The   state of the model corresponds to the vector of current molecular counts,
i.e., the number of molecules of each chemical species, and
chemical reactions trigger state transitions by changing the molecular populations. 
A system of ordinary differential equations, the
chemical master equation~\cite{mcquarrie}, describes the 
evolution of the state probabilities over time.

A classical maximum likelihood (ML)  approach, in which the likelihood
is directly approximated, is possible if all 
   populations are small~\cite{MLEJournal} or if the   model shows 
   simple dynamics (e.g. multi-dimensional normal distribution
  with time-dependent mean and covariance matrix) such that the likelihood 
  can be approximated by a normal distribution \cite{milner2013moment}.
  In this case, the likelihood (and its derivatives) can usually be approximated
  efficiently and global optimization techniques are employed to find  
   parameters that maximize the likelihood.
However, if   large populations are present in the system  then 
direct approximations of the likelihood are unfeasible since
the underlying system of differential equations contains one equation
for each state 
 and the main part of the
probability mass of the model distributes on an intractably large number
of states.
Similarly,   
if  the system shows complex dynamics such as multimodality,
  approximations of the likelihood based on
Gaussian distributions become inaccurate.

In the last years several  methods have been developed to accurately
simulate the  moments of the underlying probability distribution up
to a certain order $k$ over time~\cite{singh2006lognormal,StumpfJournal,Engblom}. 
The complexity of these simulation methods is therefore independent of the population sizes but, for large $k$, the corresponding
differential equations may become stiff and lead to poor approximations.
However, reconstructions of complex distributions from their moments show that 
for many systems 
already for small $k$ (e.g. $k\in \{4,\ldots,8\}$) the moments contain sufficient
information about the distribution such as the strength and location of 
regions of attraction (i.e. regions of the state space containing a large proportion of the probability mass)~\cite{Andreychenko2015}.

For models with complex distributions such as multiple modes or oscillations, the accuracy and the running time of the moment approximation can be markedly improved, when conditional moments are considered in combination with the probabilities
of appropriately chosen system modes such as the activity state of the genes in a gene regulatory network
~\cite{CMSB10,menz2012hybrid,Hellander2007100,jahnke2011reduced}.
Recently a full derivation of the conditional moment equations
was derived  and numerical results show that when the maximum order of the considered moments is high, the number of equations that have to be integrated
is usually much smaller for the conditional moments approach and   the resulting equations are less stiff~\cite{Hasenauer2013}. 
In addition, the approximated (unconditional) moments are more accurate when the same maximal order is considered.

An obvious parameter inference approach is the matching of the observed sample
moments with those of the moment-based simulation of the model.
Defining the differences between  sample and   (approximated) population moments
as cost functions that depend on the parameters, an approach that  minimizes the sum of the squared cost functions seems reasonable.  
  However, in a simple least-squares approach low moments such as
  means and (co-)vari\-ances contribute equally to the sum of squared differences
  as higher moments, whose absolute magnitudes are much
  higher (even if they are centralized). 
  Moreover, correlations between the different cost functions may exist and thus 
  necessitate an approach where also   products of two different cost functions 
  are considered. 
  
  The generalized method of moments (GMM) that is widely used in econometrics
  provides an estimator that is computed after assigning appropriate weights
  to the different cost function products~\cite{hansen1982large}.
  The GMM estimator has, similar as   the ML estimator, desirable statistical 
  properties such as being consistent and asymptotically normally distributed.
  Moreover, for optimally chosen weights it is an asymptotically efficient 
  estimator, which implies that (asymptotically) it has minimum variance among all estimators for the unknown parameters. 

In this paper we explore the usefulness of the GMM for moment-based
simulations of stochastic reaction networks.
We focus on two particular estimators that are commonly used in econometrics:
the two-step estimator of Hansen~\cite{hansen1982large} and the demean estimator~\cite{hall2005generalized}.
We  study the accuracy and   variance of the estimator  for different maximal moment orders
and different sample sizes by applying the GMM to two case studies.
In addition, we show that  poor approximations of some higher order moments
have a strong influence on the quality of the estimation.
Interestingly, we see that the additional information about the covariances 
of the cost functions can lead to identification of all parameters. 
In addition, the variance of the estimator
becomes smaller when higher order moments are included.
Compared to the simple least-squares approach, the GMM approach
yields very accurate estimates.

\section{Methods}
 \subsection{Stochastic chemical kinetics}\label{sec:sck}
Our inference approach relies on a Markov modeling approach that follows 
   Gillespie's theory of stochastic chemical kinetics. 
We consider a well-stirred mixture of $n$ molecular species in a volume 
 with fixed size and fixed temperature and represent it as 
 a discrete-state Markov process  
 $\left\{\bfX(t),t\ge 0\right\}$ in continuous-time~\cite{gillespie77}. 
 The random vector $\bfX(t)=\left(X_1(t),\ldots,X_n(t) \right)$ 
 describes the chemical populations at time $t$, i.e., $X_i(t)$ is 
 the   number of molecules of type $i\in\{1,\ldots,n\}$ at time 
$t$.
 Thus, the state space of $\bfX$ is $\mathbb Z^n_+=\{0,1,\ldots\}^n$.
The state changes of $\bfX$ are triggered by the occurrences of
chemical reactions. Each of the $m$ different reaction types has an associated  
non-zero change vector $\chv_j\in\mathbb Z^n$ ($j\in\{1,\ldots,m\}$), where 
$\chv_j=\chv_j^-+\chv_j^+$ such that $\chv_j^-$ ($\chv_j^+$) contains only 
non-positive (non-negative) entries and specifies how many molecules of each
species are consumed (produced)  if an instance of the 
reaction occurs, respectively.
Thus, if $\bfX(t)=\bfx$ for some $\bfx\in\mathbb Z^n_+$ with 
$\bfx+\chv_j^-$ being non-negative, then $\bfX(t+dt)=\bfx+\chv_j$
is the state of the system after the occurrence of the $j$-th 
reaction within the infinitesimal time interval $[t,t+dt)$.
W.l.o.g. we assume here that all vectors $\chv_j$
are distinct. 

We use
$\alpha_1,\ldots,\alpha_m$ to denote the propensity functions of the reactions, where 
$\alpha_j(\bfx)\cdot dt$ is the probability that, given $\bfX_t=\bfx$, 
one instance of the $j$-th reaction occurs within $[t,t+dt)$.
Assuming law of mass action kinetics, $\alpha_j(\bfx)$ is chosen 
proportional to the number of distinct
reactant combinations in state $\bfx$. 
An example is given in Table~\ref{tab:ex1},
where the first reaction gives as change vectors, for instance, $\chv_1^- = (-1,0,0)$, $\chv_1^+ = (0,1,0)$,
$\chv_1=(-1,1,0)$.  Note that, given the initial state 
$\bfx=(1,0,0)$, at any time either the DNA is active or not, i.e.
$x_1 = 0$ and $x_2=1$, or $x_1=1$ and $x_2=0$. Moreover, 
the state space of the model is infinite in the third dimension.
Although our inference approach can be used for any model parameter 
in the sequel we simply assume that
  the proportionality constants $c_j$ are  
 unknown and have to be estimated based on experimental data.
 

 For $\bfx\in \mathbb Z^n_+$ and $t\ge 0$, let $p_t(\bfx)$ denote the 
probability $P\left(\bfX(t)=\bfx\right)$. Assuming fixed initial conditions $p_0$
 the  evolution of $p_t(\bfx)$    is given by the 
 chemical master equation (CME)~\cite{mcquarrie}
$$
 \begin{array}{r@{\ }c@{\ }l}
 \frac{\partial}{\partial t}p_t(\bfx)=\!\!\sum\limits_{j:\bfx\!-\!\chv_j^-\ge 0}\! \alpha_j(\bfx\!-\!\chv_j) 
p_t(\bfx\!-\!\chv_j)-\alpha_j(\bfx)
p_t(\bfx),
  \end{array}
$$
which is an ordinary first-order differential equation  that has a unique solution under certain mild regularity
conditions.  Since for realistic systems the number of states is 
very large or even infinite, applying standard numerical solution techniques  to the CME
is infeasible. 
 If  the populations of all species remain
 small (at most a few hundreds) 
 then  the CME can be efficiently
  approximated using projection methods~\cite{sliding, Munsky06}
 or fast uniformization methods~\cite{FAUIET,Inexact}.
Otherwise, i.e., if the system contains large populations, then
analysis methods with running times independent of the population sizes
have to be used such as moment closure approaches~\cite{singh2006lognormal, StumpfJournal, Engblom} or methods
based on van Kampen's system size expansion~\cite{van1992stochastic, thomas2015}.
For both approaches, accurate reconstructions of the underlying
probability distribution, i.e., the solution of
the CME, are possible~\cite{Andreychenko2015, thomas2015}.

\begin{table}[t]
\caption{Simple gene expression model~\cite{Timmer}: The  evolution of the molecular populations \label{tab:ex1}
 DNA$_{\text{ON}}$, DNA$_{\text{OFF}}$, and mRNA is described by the random vector $\bfX(t)\!=\!(X_1(t),\!$ $X_2(t),\!X_3(t))$, respectively.}
 \begin{tabular}{lll}\hline\\[-2ex]
Reactions & Propensities & Intervals \\ 
\hline  \\[-2ex]
DNA$_{\text{ON}}\!\to\!$ DNA$_{\text{OFF}}$ & $\alpha_1(\bfx) = b\!\cdot\!x_1$ & $b\in\left[0, 0.5\right]$
\\ 
DNA$_{\text{OFF}}\!\to\!$ DNA$_{\text{ON}}$ &  $\alpha_2(\bfx) = a\!\cdot\!x_2$& $a\in\left[0, 0.5\right]$\\
DNA$_{\text{ON}} \!\to\!$ DNA$_{\text{ON}} $  & 
 $\alpha_3(\bfx) = c\!\cdot\!x_1$& $c\in\left[0, 0.5\right]$ \\
 ~~~~~~~~~~~~~~$+$ mRNA &~&~\\[1ex]\hline
\end{tabular}
\end{table}

 \subsection{Moment-based Analysis}\label{sec:mm} 

From the CME it is straightforward to derive the following equation for the derivative of the
mean of a polynomial  
function $T: \mathbb Z^n_+ \to \mathbb R$ on ${\bf X}(t)$. 
\begin{equation}\label{eq:expgen1}
\begin{array}{r@{\,}l}
 & \frac{d}{dt}E[T{({\bf X}(t))}]  \\[1ex]
=& \sum\limits_{{j}=1}^{m}\! E\left[\alpha_{j}({\bf X}(t))\! \cdot\!  \left(T({\bf X}(t) +  v_{j})  -  T({\bf X}(t)) \right)\right]
\end{array}
\end{equation}
Omitting the argument $t$ of ${\bf X}$ and 
 choosing $T({\bf X})=X_i, X_i^2,\ldots$ yields the following equations for the 
 (exact)  time evolution of the $k$-th  moment $E[{X}_i^k]$ of the distribution for the $i$-th 
species.
\begin{equation}\label{eq:expgen2}
\begin{array}{r@{\,}l}
&  \frac{d}{dt}E[{({X}_i)^k}]  \\[1ex]
 =& \sum\limits_{{j}=1}^{m} E[{\alpha_{j}({\bf X}) \cdot \left(  ({X}_i +  v_{ji})^k -  ({X}_i)^k \right)}],
\end{array}
\end{equation}
where $v_{ji}$ refers to the $i$-th component of the change vector $\chv_j$.
In a similar way, equations for mixed moments are derived.

If all reactions are at most monomolecular ($1\ge \sum_i |v_{ji}^-|$ for all $j$), then no moments of order higher than $k$ appear 
on the right side (also in the mixed case) and we can directly integrate all equations for moments
of at most order $k$. 
However,   most systems \emph{do} contain bimolecular reactions (in particular those
with complex behavior such as multistability). 
 In this case we consider a 
 Taylor expansion of the multivariate function 
 $$
 f({\bf X})={\alpha_{j}({\bf X}) \cdot \left(T({\bf X} +  v_{j}) -  T({\bf X}) \right)}
 $$
  about the mean $\mu:=E[{{\bf X}}]$. 
  It is easy to verify that, when applying the expectation to the Taylor sum, the right side
  only contains derivatives of $f$ at   ${\bf X}=\mu$,   which are multiplied 
  by  central moments of increasing order.
  For instance, for $k=1$ and a  single species system with $n=1$, Eq. \eqref{eq:expgen2}  becomes
  $$
  \begin{array}{lcl}
    \frac{d}{dt}E[{({X}_i)}] 
  &=& \sum\limits_{{j}=1}^{m}   v_{ji} E[\alpha_j({\bf X})]\\[1ex]
  &=& \sum\limits_{{j}=1}^{m}   v_{ji} \big(\alpha_j(\mu)  +\frac{E[({\bf X}-\mu)]}{1!} \cdot \frac{\partial}{\partial x}\alpha_j(\mu) \\[1.5ex]
  && ~~~~
 +\frac{E[({\bf X}-\mu)^2]}{2!} \cdot \frac{\partial^2}{\partial x^2}\alpha_j(\mu)+\ldots\big)
  \end{array}
  $$
  In the expansion, central moments of higher order may occur. For instance, in the case of bimolecular 
  reactions, the equations for order $k$ moments involve central moments of order $k+1$ since second order
  derivatives are non-zero. 
By converting the non-central moments to central ones and truncating
  the expansion at some fixed maximal order $k$, we can close the system of equations when we assume that higher order central moments are zero.   
A full derivation of the   moment equations using multi-index notation (as required for $n>1$) can be
found in \cite{Engblom}. 

The accuracy of the inference approach that we propose in the sequel depends not only on the information given by
the experimental data but also on the accuracy of the approximated moments. 
Different closure strategies have been suggested and compared in the last years
showing that the accuracy can be improved by making assumptions about the 
underlying distribution (e.g. approximate log-normality) \cite{schnoerr2015, bogomolov2015adaptive}.
In addition, the accuracy of moment-closure approximations
has been theoretically investigated~\cite{grima2012study}.


 \subsection{Hybrid Approaches}\label{sec:hybrid}

Compared to deterministic models that describe only   average behaviors, stochastic models provide interesting additional information about the behavior of a system.
Although this comes with additional computational costs, it 
is in particular for systems with complex   behavior, such as 
multimodality or oscillations, of great importance. 
Often the underlying source of multiple modes are discrete
changes of gene activation states that are described by 
chemical species whose maximal count is very small (e.g. 1 for the
case that the gene is either active, state 1, or inactive, state 0). Then the   moment-based approaches described above can be 
improved (both in terms of accuracy and computation time) by
considering conditional moments instead \cite{CMSB10,menz2012hybrid,Hellander2007100,Hasenauer2013,Soltani2015}.
The idea is to split the set of species into species with
small and large populations and consider the moments of the
large populations conditioned on the current count of the 
small populations.
For the small populations, a small master
equation has to be solved additionally to the moment equations
to determine the corresponding discrete distribution.
More specifically, if $\hat{\bfx}$ is the subvector of $\bfx$
that describes the small populations and $\tilde\bfx$ is the 
subvector of the large populations (i.e. $\bfx=(\hat{\bfx},\tilde{\bfx})$), then for the distribution of
 $\hat{\bfx}$ we have
 $$
 \begin{array}{r@{\,}c@{\,}l}
\frac{d}{dt} p_t(\hat{\bfx}) &=& \sum\limits_{j:\hat{\bfx}-\hat{\chv}_j\ge 0} \!\!\! E[\alpha_j({\bf X})\mid \hat{\bf X}=\hat{\bfx}-\hat{\chv}_j] p_t(\hat{\bfx}-\hat{\chv}_j) \\[1.5ex] && -  \sum_{j}  E[\alpha_j({\bf X})\mid \hat{\bf X}=\hat{\bfx}] p_t(\hat{\bfx}) 
 \end{array}$$
 where $\hat{\chv}_j$ is the corresponding subvector of ${\chv}_j$. Using Taylor expansion, the conditional expectations of the propensities can, as above, be expressed in terms of conditional moments
 of the large populations. In addition, equations for the 
 conditional moments   of the large populations
 can be derived in a similar way as above. For instance, the partial mean
 $E[\tilde X_i \mid \hat x] p_t(\hat x)$ follows the time evolution
 $$ \begin{array}{l}
 \frac{\partial}{\partial t} \left(E[\tilde X_i \mid \hat x] p_t(\hat x) \right) \\[1.5ex]
   \ =\!\!\! \sum\limits_{j:\hat{\bfx}-\hat{\chv}_j\ge 0}\!\!\!\!\!\!  E[(\tilde X_i+v_{ij})\alpha_j({\bf X})\mid \hat{\bf X}=\hat{\bfx}-\hat{\chv}_j] p_t(\hat{\bfx}-\hat{\chv}_j) \\[2.5ex]  \ -  \sum_{j}  E[\tilde X_i\alpha_j({\bf X})\mid \hat{\bf X}=\hat{\bfx}] p_t(\hat{\bfx}) 
 \end{array}$$
where on the right side again Taylor expansion can be used to replace unknown conditional expectations by conditional moments. As above a dependence on higher conditional moments may arise and a closure approach has to be applied 
to arrive at a finite system of equations.
 Unconditional
moments can then be derived by summing up the weighted 
conditional moments.
 It is important to note that if $p_t(\hat x)=0$ then algebraic equations arise turning the
 equation system into a system of differential-algebraic equations,
 which renders its solution   more difficult (see \cite{Hasenauer2013,kazeroonian2016cerena} for details). 
 
\begin{figure}[t]
\includegraphics[width=0.5\textwidth]{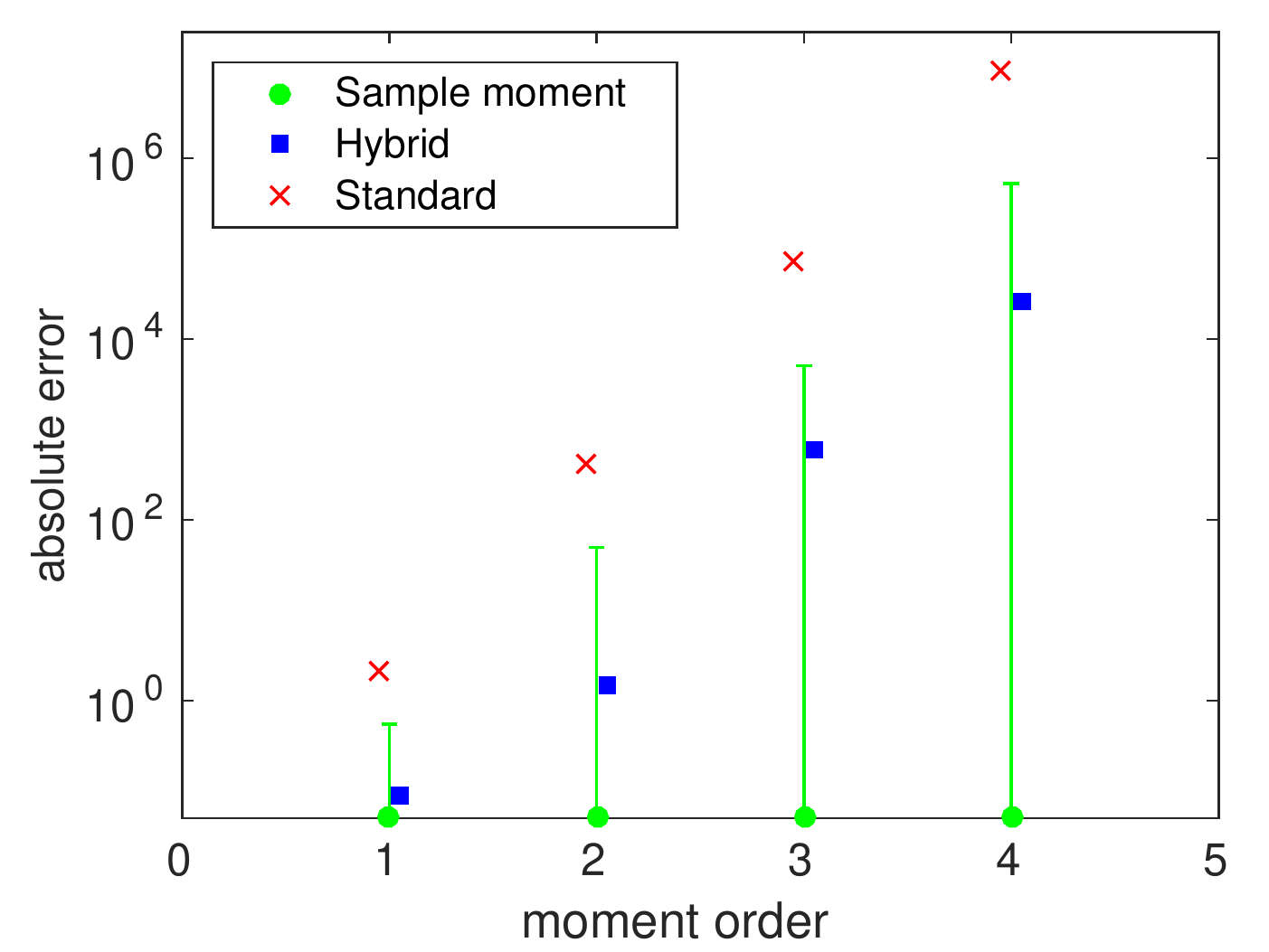}
\caption{Absolute error of the first four moments of P$_1$ for the exclusive switch model, where the moments are either computed based on a standard moment closure approach or a hybrid approach. The maximal order of the considered moments is $5$. }\label{fig:MomComp}
\end{figure}

 In Fig.~\ref{fig:MomComp} we give an example for a comparison of the accuracy of the hybrid approach 
and the standard moment closure (assuming that all central moments above a fixed maximal order
are zero) for one of our case studies. As ''exact'' moment values we chose the average
of 500,000 samples generated by the stochastic simulation
algorithm (SSA)~\cite{gillespie76} and considered the absolute difference to the approximated moments of
one chemical population until a maximal order of four. Since for our case studies we assumed
10,000 samples we additionally plot the (approximated) standard deviation of the 50 sample means 
taken from batches of 10,000 samples. The moments computed based on the hybrid approach 
show a smaller error than those computed using the standard moment closure and lie
within the deviations given by the sample moments.
For the example in Fig.~\ref{fig:MomComp} we
have 126 equations for the standard approach up to an order of four. 
In the hybrid case there are 14 moment equations and one equation for the mode probability per mode leading to a total number of 45 equations.
However, reductions are possible for the standard approach when the model structure is exploited \cite{ruess2015minimal}.
 We do not make use of these reductions here but choose the hybrid approach mainly because it gives more accurate results for the (unconditional) moments.
 This strongly improves 
 the quality of the estimated parameters 
 as demonstrated in the Results section.

 \subsection{Generalized Method of Moments}\label{sec:gmm}
 
We assume that  observations
of a biochemical network were made
using single-cell analysis 
that gives population snapshot data
 (e.g. from flow cytometry measurements).
Typically, large numbers (about 5,000-10,000 \cite{HanleyLomasMittarMainoPark2013,Hasenauer2011,Munsky201512}) of independent samples 
can be obtained where each sample corresponds to one cell.
It is possible to simultaneously observe one or several chemical populations  
at a time in each single cell. In the sequel, we first describe the inference procedure for a single observation
time point and a single chemical species that is observed. 
 Later, we extend this to several time points and species.

For  a fixed measurement time $t$ and a fixed index $i$ of the observed population  we can define the   $r$-th order sample moment as
 $$
 \hat m_r = \frac{1}{N} \sum_{\ell=1}^N Y_{\ell}^r,
 $$
where $Y_{\ell}$ is the $\ell$-th sample of the observed molecular count 
of the $i$-th species  at time $t$ and there are $N$ samples in total. 
For large $N$, the sample moments are asymptotically unbiased
estimators of the population moments.

Let $\theta$ be a vector of, say, $q\le m$ unknown reaction rate constants\footnote{It is straightforward to adapt the approach that we present in the sequel to
the case that  other unknown continuous parameters   have to be estimated.}, for which some biologically relevant range is known.
Moreover, let $m_r$ be the $r$-th theoretical moment, i.e., $m_r(\theta):=E[Y_\ell^r]$.
In the sequel we also simply write $Y$ instead of $Y_\ell$
whenever $Y$ appears inside the expectation operator or when
the specific index of the sample is not relevant.
An obvious inference approach would be to consider the ordinary least
squares estimator
\begin{eqnarray}\label{eq:ls}
\hat{\theta} &=&\arg\min_\theta \sum_{r=1}^k \left(\hat m_r-m_r(\theta)\right)^2,
\end{eqnarray}
where $k$ is the number of moment constraints.
Under certain conditions related to the identification of the parameters as discussed below, this estimator is consistent (converges in probability to the
 true value of $\theta$)
  and asymptotically normal. However, its variance may be very high. 
  This is due to the fact that for increasing order  the variance of  the sample moments  increases and so does the variance of the estimator.
  This problem can be mitigated by choosing appropriate weights
  for the summands in~\eqref{eq:ls}. Moreover, since correlations
  between the cost functions  $$g_r(\theta)=\hat m_r-m_r(\theta)$$
  exist, a more general approach that considers mixed terms is needed.
  This leads to a class of estimators, called generalized method of moments (GMM) estimators that have been introduced by Hansen~\cite{hansen1982large}. 
The idea is to define the estimator as
\begin{eqnarray}\label{eq:gmm}
\hat{\theta} &=&\arg\min_\theta  {\bf g}(\theta)' W {\bf g}(\theta)
\end{eqnarray} 
where ${\bf g}(\theta)$ is the column vector with entries $g_r(\theta)$, $r=1,\ldots,k$,
and $W$ is a positive semi-definite weighting matrix.
Note that by defining $f_{r}(Y,\theta)=Y^r-m_r(\theta)$ we see that 
$$g_r(\theta)=\frac{1}{N}\sum_\ell f_{r}(Y_\ell,\theta)=\frac{1}{N}\sum_\ell Y_\ell^r -m_r(\theta)$$ is the sample counterpart
of the expectation $E[f_{r}(Y,\theta)]$.
The latter satisfies
  $$
  \theta_0= \arg\min_\theta  E[{\bf f}(Y,\theta)]' W E[{\bf f}(Y,\theta)],
 $$
 where ${\bf f}(Y,\theta)$ is the column vector with entries $f_{r}(Y,\theta)$
 and $\theta_0$ is the true value of $\theta$.
 Note that the choice   $W=I$ gives the least-squares
 estimator with $k$ terms while for general $W$ there are $\frac{k\cdot(k+1)}{2}$ terms in the 
 objective function (with $k$ being the dimension
 of ${\bf g}(\theta)$).
 In addition, we remark that in general $W$ may depend on $\theta$ and/or the samples $Y_\ell$.
 
 Here we assume that identification of $\theta$ is possible, 
 i.e., we require that $q\le k$, i.e., the  {number} of the moment constraints  {used}
  is at least as large as the number of unknown parameters  and $$
 E[{\bf f}(Y,\theta)]={\bf 0} \mbox{ if and only if } \theta=\theta_0.
 $$
 In addition, 
    the  theoretical moments $m_r(\theta)$ should not be functionally dependent (see Chapter 3.3 in~\cite{hall2004generalized}) 
    to  ensure that
   the  information contained
  in the moment conditions is sufficient for successfully identifying the parameters.

By applying the central limit theorem to the sample moments,
it is possible to show that the GMM estimator is consistent and
asymptotically normally distributed and that its variance
becomes asymptotically minimal if  the matrix $W$ is chosen 
such that it is proportional to the inverse  of the   
covariances between the $Y_\ell^r$~\cite{hansen1982large}. This 
result is intuitive since usually higher moments might be
more volatile than others and, thus, it makes sense to 
normalize the errors in the
moments by the corresponding covariance. Formally, we define
${\bf Y}_\ell$ as the random vector with entries $(Y_\ell)^r$
for $r=1,\ldots,k$ and, as before, omit the subindex $\ell$ if it is not relevant. Then,
$$
F(\theta_0)=\mathit{COV}[{\bf Y},{\bf Y}]=E[{\bf f}({Y},\theta_0){\bf f}({Y},\theta_0)^T]
$$
and choosing $W\propto F^{-1}$ will give an estimator with  smallest possible variance, i.e., it is asymptotically efficient in this class of estimators~\cite{hansen1982large,hall2004generalized}.
 
 
 Since $F$ depends on the true value $\theta_0$, a two-step updating procedure has been suggested \cite{hansen1982large} during which   $W$ is chosen as the identity matrix $I$ in the first step such that an initial estimate $\tilde\theta$ is computed. In a second step, $F$ is estimated by the sample counterpart of   $E[{\bf f}({Y},\tilde\theta){\bf f}({Y},\tilde\theta)^T],
 $ i.e.,\begin{equation}\label{hatF2}
 \hat F_1(\tilde\theta)=\frac{1}{N}\sum^N_{\ell=1}{\bf f}({Y}_\ell,\tilde\theta){\bf f}({Y}_\ell,\tilde\theta)^T.
 \end{equation}
If, however, the model is ``misspecified'', i.e., there is no $\theta_0$ for which
 $$E[{\bf f}({Y},\theta_0)] = \bf 0,$$ then the above estimator is no longer consistent. 
 In particular, if the theoretical moments are poorly approximated, it is likely that also
the accuracy of the resulting estimates is poor.
 An estimator for $F$ that is consistent is then given by~\cite{hall2004generalized}
\begin{equation}\label{hatF}
 \hat F_2=\frac{1}{N}\sum^N_{\ell=1} ({\bf Y}_\ell - \overline{\bf Y})({\bf Y}_\ell - \overline{\bf Y})^T,
 \end{equation} 
 where $ \overline{\bf Y}$ is the vector with entries $\frac{1}{N}\sum_{\ell=1}^N {\bf Y}_\ell^r$.
 In the sequel we refer to the estimator based on~\eqref{hatF} as the \emph{demean estimator}. 
 This estimator removes the inconsistencies in the      covariance matrices estimated from the sample moments by ''demeaning``. 
 Since moment-based analysis methods usually give   approximations of the moments and not the exact values,
 we consider both, the demean estimator defined by \eqref{hatF} and the estimator of  the 2-step procedure in \eqref{hatF2} for our numerical results.

The estimation procedure described above can be generalized 
to several dimensions by also using mixed sample moments instead of only $\hat m_r $ and mixed theoretical moments instead of only $m_r(\theta)$. For instance, for  moments up to order two and 
two simultaneously observed species $X$ and $Y$, we use 
the cost functions
$$
\begin{array}{rcl}
g_1(\theta)&=&\frac{1}{N}\sum_{\ell=1}^N X_\ell - E[X_\ell\mid\theta] \\
g_2(\theta)&=&\frac{1}{N}\sum_{\ell=1}^N Y_\ell - E[Y_\ell\mid\theta] \\
g_3(\theta)&=&\frac{1}{N}\sum_{\ell=1}^N X_\ell Y_\ell - E[X_\ell Y_\ell\mid\theta] \\
g_4(\theta)&=&\frac{1}{N}\sum_{\ell=1}^N X_\ell^2 - E[X_\ell^2\mid\theta] \\
g_5(\theta)&=&\frac{1}{N}\sum_{\ell=1}^N Y_\ell^2 - E[Y_\ell^2\mid\theta]. \\
\end{array}$$
In the same way, we can extend the estimators $\hat F_1$ and $\hat F_2$ to
several dimensions. For instance, the covariance between $X_\ell Y_\ell$ and $X_\ell^2$ can be estimated as 
$$
\frac{1}{N}\sum_{\ell=1}^N (X_\ell Y_\ell  - \overline{XY})(X_\ell^2 - \overline{X^2}),
$$
where again we use $\overline{*}$ to denote the sample mean operator. 

If, instead of snapshot data for a single observation time, independent samples for different times are available then the GMM estimator can also be easily generalized to 
\begin{eqnarray}\label{eq:gmm_time}
\hat{\theta} &=&\arg\min_\theta \sum_{t=t_0}^{t_f} {\bf g}^{\left(t\right)}(\theta)' W^{\left(t\right)} {\bf g}^{\left(t\right)}(\theta).
\end{eqnarray}
Here, for each time point $t\in\{t_0,\ldots,t_f\}$ the vector of
cost functions ${\bf g}^{\left(t\right)}$ is calculated as before and the minimum is taken over the sum of these uncorrelated cost functions. Note that for each observation time point a 
weight matrix $ W^{\left(t\right)} $ has to be computed.
In the two-step approach, the initial weight matrices are all 
equal to the identity matrix and then in the second step
different weight matrices may arise since the estimator of $F$
depends on $Y$, which in turn depends on the distribution
of the model at the specific time~$t$.

\section{Results}\label{sec:results}
  
To analyze the performance of the GMM we consider two case studies, the simple gene expression model 
in Table~\ref{tab:ex1} and a network of two genes with mutual repression, called exclusive switch \cite{loinger-lipshtat-balaban-biham07}.
The  reactions of the exclusive switch are listed in Table~\ref{tab:ex2}. 
All propensities follow the law of mass action.
For the parameters that we chose, the corresponding probability distribution
is bi-modal.

For fixed reaction rate constants and initial conditions, we 
used the SSA to generate trajectories of the systems and record samples of   the size of the corresponding protein/mRNA populations.
In addition, we used the software tool SHAVE~\cite{hscc11} to generate  moment equations both for the standard moment closure and 
for the hybrid approach. In SHAVE the partial moments are integrated instead of the
conditional moments such that the differential-algebraic equations
are transformed into a system
of (ordinary) differential equations after truncating modes
with insignificant probabilities. Then  an accurate approximation of the solution using standard
numerical integration methods can be obtained.  
The system of moment equations is always closed by setting all  central moments of order $>k$ to zero.
We used for the inference approach only the moments up to order $k-1$ since the precision of the moments
of   highest order $k$ is often poor. 
 SHAVE allows to export the (hybrid) moment equations as a MATLAB-compatible m-file.
We then used MATLAB's ode45 solver, which is based on a fifth order Runge-Kutta method, to integrate the (hybrid) moment equations.
  Note that for the gene expression example, the moment equations are exact since all
propensities are linear. Thus, even an analytic  solution is possible for this system.

 We then used MATLAB's Global Search routine to minimize the
objective function in Eq.~\eqref{eq:gmm}. 
Global Search is  a method for finding the global minimum by starting a local solver from 
multiple starting points that are chosen according to a heuristic~\cite{ugray2007scatter}.
Therefore the total running time of our
method depends on the tightness of the intervals that we
use as constraints for the unknown parameters as well as on
the starting points of the Global Search procedure. 
The running times for one local solver call (using the
hybrid approach for computing moments) were about $2~$s (demean
estimator) and $40~$s (2-Step estimator) for the gene expression model. For the exclusive switch the average running time
for a local solver call was about $2~$min (demean) and $10~$min (2-Step). Note that the total running time depends on the amount of local solver calls carried out by Global Search, which varied between 2 and 50.
For all experiments we chose a single initial point
that is located far away from the true values
and allowed Global Search to choose
500 (potential) further starting points. Different initial
points yielded similar results except if the initial
points is chosen close to the true values (then the
results are significantly better in particular in the
case of only few moment constraints).

The intervals that we used as constraints for
the parameters are all listed in Tables~\ref{tab:ex1}
and \ref{tab:ex2}.

\begin{table}[t]
\caption{Exclusive switch model~\cite{loinger-lipshtat-balaban-biham07}:   \label{tab:ex2} 
Two different proteins P$_1$ and P$_2$ can bind to a promotor region on the DNA. If P$_1$ is bound to the promotor the production of P$_2$ is inhibited and vice versa. In the free state both proteins can be produced. 
}
\begin{tabular}{r@{\,}c@{\,}lcc}\hline
\multicolumn{3}{c}{Reactions, $i=1,2$} & Constant & Interval\\ 
\hline  \\[-2ex]
DNA &$\to$ &DNA + P$_i$ & $p_i$ & [0.5,1.5]
\\ 
DNA.P$_i$&$\to$ &DNA.P$_i$ + P$_i$ &  $p_i$ & [0.5,1.5]\\
P$_i$&$\to$&$ \emptyset$ & $d_i$ & [0,0.05]\\
 DNA + P$_i$&$ \to$ &DNA.P$_i$ & $b_i$ & [0,0.1]\\
DNA.P$_i$&$ \to$& DNA + P$_i$ & $u_i$ & [0,0.1]
 \\[1ex]\hline
\end{tabular}
\end{table}

\begin{figure*}[t]
        \includegraphics[width=0.99\textwidth]{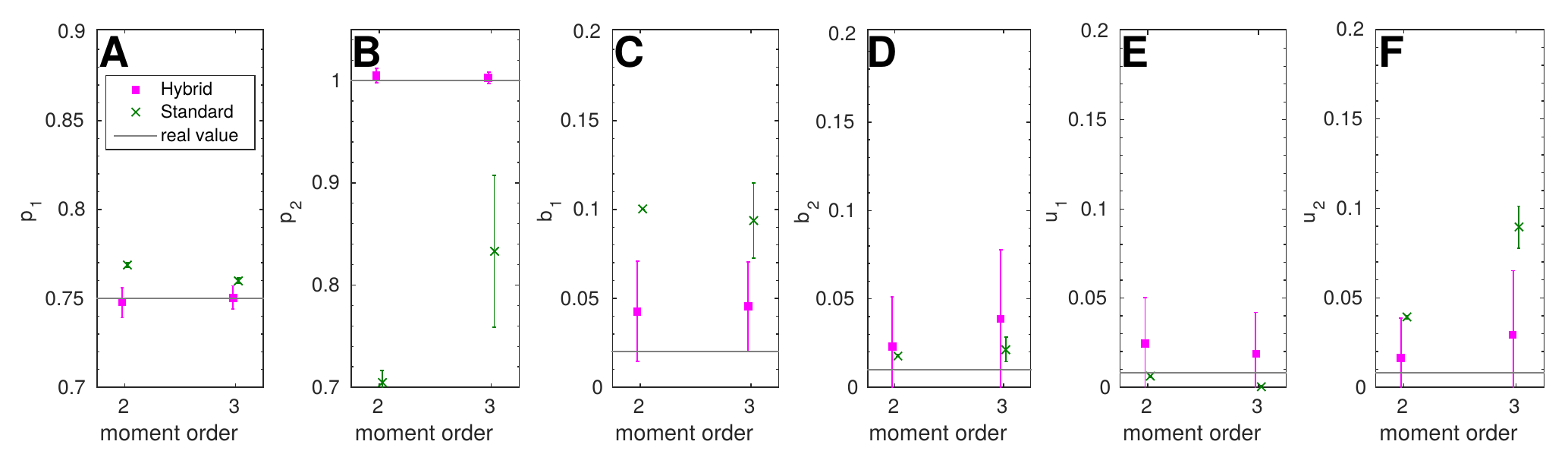}
    \caption{Exclusive switch model: Comparison of estimations with the demean procedure for the standard moment closure and hybrid moments. }\label{fig:STvsHYB}
\end{figure*}
\subsection{Standard vs. hybrid moment-based analysis}\label{sec:SVH}

 In Fig.~\ref{fig:STvsHYB} we plot the results of a comparison between 
 the standard and the hybrid moment closure when it is performed during the
 optimization procedure of the GMM inference approach.
  We chose the exclusive switch model for this since for this model the
  accuracy of the standard approach is poor. 
 As an estimator for $F$ we used~\eqref{hatF},
  which is based on demeaning (demean). Results for the 2-step procedure 
 show similar differences when standard and   hybrid moment closure are compared. 
 We fixed the degradation rates to ensure that identification of $p_1$ and $p_2$ is possible 
 when the two protein populations are measured at 
 only a single observation time point. To simultaneously identify
 all parameters (including $p_1$ and $p_2$) 
 several observation time points are necessary (see Appendix~\ref{app}).
 \newpage
 The true values of the six unknown parameters are plotted against
  the means and standard deviations of the estimated values    for a
  maximal moment order of 2 and 3,
     where for each of  the six unknown parameters 50 estimations based on 10,000 samples each were used.

 We see that the inaccurately approximated moments of the standard approach
  lead to severe problems
 in the inference approach. Nearly all parameters are estimated more accurately
 when the hybrid moment closure is used. For parameter $b_1$ most of the
  optimization runs converged to the upper limit of the given interval (0.1)
  when the standard approach was used.
  For the results in the sequel, we only used the hybrid moment closure.

\begin{figure*}[t]
        \includegraphics[width=0.99\textwidth]{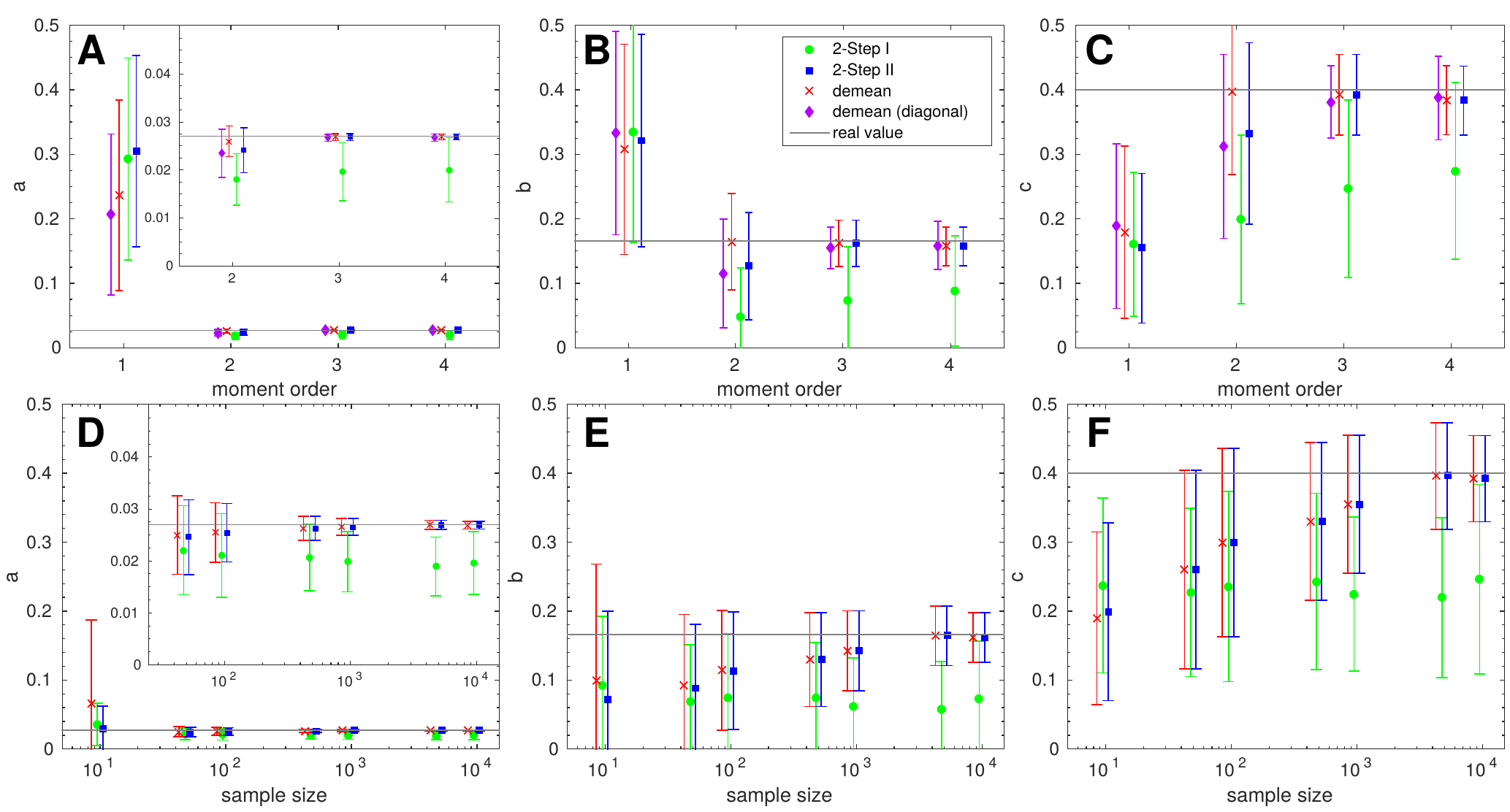}   
    \caption{Gene expression model:  Estimated parameters $a$,$b$ and $c$ for
      different numbers/orders of moments and $10{,}000$ samples (\textbf{A}-\textbf{C}) and for
    different sample sizes based on 3 moments (\textbf{D}-\textbf{F}). The inner plots 
   show results on a more detailed scale (\textbf{A} and \textbf{D}).}
    \label{fig:GenEx}
\end{figure*}

 \subsection{Two-step vs. demean approach}
\begin{figure*}[t]
        \includegraphics[width=0.99\textwidth]{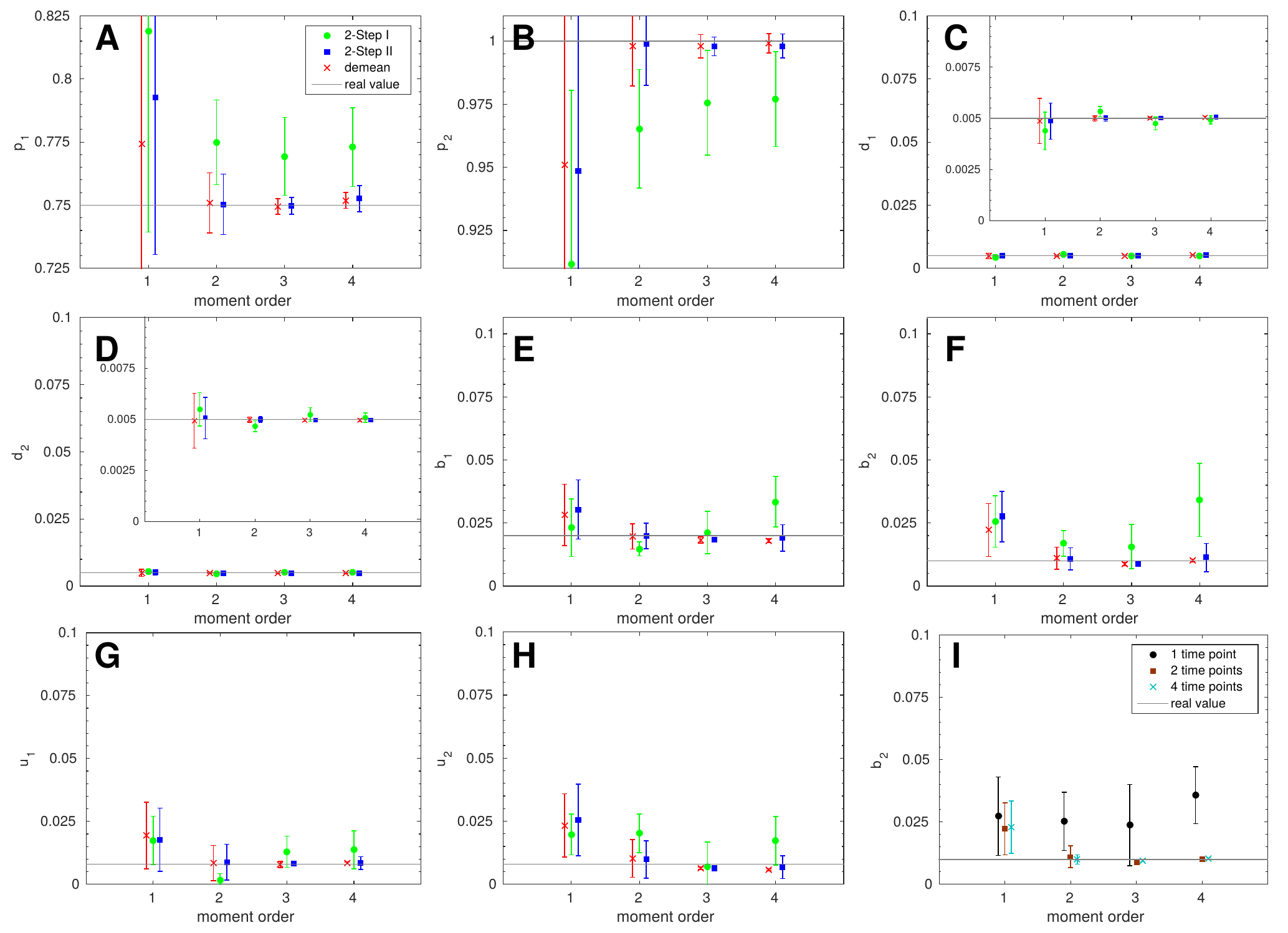}
  \caption{Exclusive switch model: Estimated parameters for maximal moment order 1-4 based on $10{,}000$ independent samples observed at time $t=100$ and $t=200$ (\textbf{A}-\textbf{H}) and at 1-4 different time points for the demean-based estimation of $b_2$ (\textbf{I}). The inner plots 
    show results on a more detailed scale (\textbf{C} and \textbf{D}).}\label{fig:EXS}
 \end{figure*}

 In Fig.~\ref{fig:GenEx} and \ref{fig:EXS} we plot results
of the GMM approach applied to the two example networks, where we compare the performance of
the two-step estimator in Eq.~\eqref{hatF2} with the demean estimator in Eq.~\eqref{hatF}.
We plot the true values of the parameters against the estimated values, where 2-Step I is the result of the 
first step of the two-step procedure (with $W=I$) and 2-Step II that of the second step (with $W=\hat F_1$ and $\hat F_1$ as defined in Eq.~\eqref{hatF2}).

  For the results in Fig.~\ref{fig:GenEx} only one population (mRNA) was observed at $t=100$ where the initial conditions were such that DNA$_{\text{OFF}}=1$,  DNA$_{\text{ON}}=0$ and 10 mRNA molecules were present in the system.
  For three parameters the means and standard deviations of the estimated  
  values are plotted, again  based on
  50 repetitions of the inference procedure. 
   
  In the first row of Fig.~\ref{fig:GenEx}
  the accuracy of the estimation is compared with respect to
  the number/order of moments considered,
  where again for each of the 50 estimated values 10,000 samples were used.
    We see that if only one  moment is considered or if equal weights
  are used for the first two moments, 
  only    a rough estimate is possible since identification is not possible.
  The  accuracy is markedly 
  improved when the weights are chosen according to the 
  demean approach. Here, it is important to note that for a maximal
  order of $k=2$, 
  in $W$ we also consider, besides the
  squared cost functions $g_1(\theta)^2$ and $g_2(\theta)^2$,  
   the mixed term $g_1(\theta)g_2(\theta)$. 
This additional term significantly improves the quality of the estimation
such that it is possible to  achieve a good estimation of the parameters with 
only the sample mean and the sample  second moment.
To further investigate the positive influence of the mixed term, we additionally plot results 
for the case that only variances are estimated, referred to as 
'demean (diagonal)', i.e., the weight matrix is the inverse of a diagonal matrix
  that  contains the    variances estimated based on the demean approach.\\
However, the variance of the estimator for a maximum order of two is relatively high but decreases 
significantly when also the third (and fourth) moment is considered.
Here, demean and the second step of the two-step procedure perform equally
well and also demean (diagonal) gives very good results. Opposed to this  $W=I$ (first step of two-step procedure) gives 
poor results and a high variance also if higher moments are considered.

 In Table~\ref{tab:weights} we give an example for the (normalized) matrix $W$ as used for demean and 
 2-Step II.
The two methods choose  nearly identical weights and the mean 
has the highest weight. Then, the mixed cost function for 
mean and second moment has a (negative) weight  of about $2\cdot (-4.95)$\% since these moments are negatively correlated (and so are the second and third moment). 
All terms that involve the third moment have a very small weight as their covariances are   high.  
 
It is important to note that also if the number of moment constraints, $k$, is equal to
the number of parameters, $q$, 2-Step I performs poor (see results for maximal order $k=3$ in the
first row of Fig.~\ref{fig:GenEx}).
The reason is that in this example  identification is not possible if only three terms are used due to functional dependencies
between the parameters of the first two reactions and due to the fact that only at a single time
point measurements were made.
If identification was possible and the computed population moments were exact, 
the results should be independent  of the choice of $W$ for the case that $q $ equals~$k$.    

 Thus, the weights given by the estimators for $F$
 in~\eqref{hatF2}
 and~\eqref{hatF} substantially increase the accuracy  of the results 
 and allow identification, because additional information about the
 covariances between the $Y^r$ are used.
 Moreover, due to   the off-diagonal entries of $W$ additional mixed terms 
 are part of the objective function.

 In the second row in Fig.~\ref{fig:GenEx}, we compare 
 the accuracy for different samples sizes where the first 
 three moments were considered. 
 While   2-Step I does not
 show a systematic improvement when the number of samples 
 increases, we see for 2-Step II and demean not only significantly improved
 estimates but also smaller variances. 
However, in the case of few samples, demean gives in particular for parameter $a$ a high variance. This
comes from the fact that the corresponding estimator
uses the sample mean instead of the theoretical mean
and therefore the weight matrix is far from optimal
if $N$ is small.
  
In Fig.~\ref{fig:EXS}, \textbf{A}-\textbf{H}, we plot results for the exclusive switch model where all eight parameters were estimated based on observations of the 
two protein populations of $P_1$ and $P_2$ at two time points.
On the x-axis the maximal order of moments used
is plotted. 
For the orders 1, 2, 3 and 4 there are in total 2, 5, 9 or 14 moments, respectively.
Again, 2-Step II and demean both give accurate results from a maximal order of two on, whereas 
2-Step I gives poor results.
In addition, the variance of the estimator decreases with increasing maximal order.
However, the values for 2-Step II become  slightly worse and have higher variance
for a maximal order of four since these moments are not approximated very accurately.
Also the accuracy of the demean estimator   does not improve when the maximum order is
increased from three to four. Thus, the cost functions of order four moments do not lead to any significant improvement
in this example and should be excluded.

\begin{table}[t]
\caption{Weight matrices for the two-step and demean procedure with moment order 3 for the gene expression model. The entries are normalized with respect to the weight for the mean and rounded (the original weight matrices are both positive semi-definite).}
\begin{tabular}{lc}\hline\\[-2ex]
~& $W$ \\ \hline  
 Two-Step &
  $  
  \begin{array}{ccc}
  & & \\[-1ex]
  1    &   -0.0495  &    0.0007\\
      -0.0495  &    ~~0.0025 &    -3.86e^{-5}\\
      ~~0.0007  &  ~-3.86e^{-5}  &    ~~6.11e^{-7}
 \end{array}  $\\
 \hline 
  Demean &
 $  
 \begin{array}{ccc}
  & & \\[-1ex]
1    &   -0.0494   &   0.0007\\
      -0.0494  &    ~~0.0025   &  -3.85e^{-5}\\
       ~~0.0007 &   ~-3.85e^{-5}  &    ~~6.09e^{-7} 
 \end{array}  $ \\
 \hline 
\end{tabular}
\label{tab:weights}
\end{table}

 \subsection{Further  estimators}
For our results we focused on the most popular GMM estimators, that is, demean and two-step.
However, we also implemented two additional variants of estimators that are frequently described in the 
GMM literature \cite{hall2005generalized,hansen1996finite}. One is the estimator that results from further iterations of the two-step procedure
(iterated GMM estimator \cite{hansen1996finite}).
However, in our examples we did not see an increase in accuracy after the second iteration. 
 The second approach is the continuously
updating GMM estimator \cite{hansen1996finite}, where 
we use in Eq.~\eqref{eq:gmm} the  weight matrix $W(\theta)=(\hat{F}_1(\theta))^{-1}$ of Eq.~\eqref{hatF2}  
 and the argument $\theta$ is not fixed for the optimization but optimized simultaneously with the
argument of ${\bf g}(\theta)$. The results for this approach did not show increased
accuracy, also  when we used results of the other estimators (e.g. demean) 
as starting points for the optimization. Moreover, for large weight matrices, the recomputation in each
step of the optimization resulted in longer running times.\\
Overall, our experiments show that for sufficiently large  $N$ the demean estimator usually
yields the best results, while two-step performs better for small $N$.
Moreover, choosing three as the maximum order gave the best results (accurate average value
and small standard deviations) for the examples that we considered.
\section{Discussions} \label{sec:related}
 In the context of stochastic chemical kinetics, 
parameter inference methods are either based on 
  Markov chain Monte Carlo schemes ~\cite{Wilkinson2,wilkinson2011stochastic,golightly2011bayesian,daigle2012accelerated}, on approximate Bayesian computation   techniques 
  \cite{drovandi2011estimation,fearnhead2012constructing,Stumpf}
or on maximum likelihood estimation using a direct approximation of the likelihood~\cite{MLEJournal,Timmer} or a simulation-based estimate \cite{Tian2007,Uz2010478}. 
Maximum likelihood
estimators are, in a sense, the most 
informative estimates of   unknown parameters~\cite{Higgins}
and have desirable mathematical properties such as
    unbiasedness,
   efficiency, and    normality. 
On the other hand, the computational complexity of 
maximum likelihood estimation is high as it requires
a simulation-based or numerical solution of the CME for many different parameter
instances. Since the applicability of these
 methods is limited,  approaches based on moment closure~\cite{milner2013moment,bogomolov2015adaptive,zechner2012moment,ruess2015moment,kugler2012moment,Hasenauer2016}  or linear
 noise approximations~\cite{Komorowski2009,zimmer2015deterministic,bergmann2016piecewise} have been
 developed.  
  An approximation of the likelihood of   order-two sample
moments is maximized in \cite{bogomolov2015adaptive,zechner2012moment,ruess2015moment,Hasenauer2016}. 
The approach exploits that for large numbers of samples these 
sample moments are asymptotically normally distributed. 
The negative log-likelihood   leads to an optimization problem 
where the differences between the sample and theoretical moments up to order two are weighted and minimized as well. 
 As opposed to the GMM, the weight matrix in \cite{zechner2012moment,ruess2015moment} is  
 estimated   based on the theoretical moments of the model
up to order four and independent of the samples while in the GMM approach this matrix  depends on the 
samples (and theoretical moments up to order two). Moreover, the objective function contains an additional summand, which is the logarithm of the determinant
of the estimated covariance matrix. 
In \cite{bogomolov2015adaptive},  Bogomolov et al.
 insert   sample instead of   theoretical moments 
 in the derived formulas for the covariances of moment conditions up to order two. 
 A comparison for the two examples that we consider in the previous section yields that  when the theoretical moments are used
 to estimate covariances, 
 similar to the continuously updating GMM,   optimization was slow and sometimes failed
   to return the global optimum due to a much more complex landscape of the objective function.
   When sample moments are considered as suggested in \cite{bogomolov2015adaptive},
 the results are similar to those of the GMM demean estimator for a maximum order of two.
 In \cite{Hasenauer2016}, only variances are considered (weight matrix is diagonal) and estimated 
based on the samples. Therefore, it does not exploit the information contained in the mixed terms,
which lead to improved estimates in our examples (see results for  'demean (diagonal)' in Fig.~\ref{fig:GenEx}).

A similar approach is used in \cite{milner2013moment} where the moment equations are closed by a Gaussian approximation. The parameter estimation is based on using a ML estimator and a Markov chain Monte-Carlo approach.
In \cite{kugler2012moment} the importance of  higher moment orders when using least square estimators is shown. 
Weights for terms that correspond to different moments are chosen ad hoc and not based on any statistical framework.

Here, we present results for the general method of moments that assigns optimal weights to the different 
moment conditions for an  arbitrary maximal moment order and number of species. We showed that trivial weights (e.g. identity matrix)
give   results whose accuracy  can be strongly increased when optimal
weights are chosen. In the very common case that functional dependencies between
parameters exist (e.g. degradation and production of the same species) 
and identification is difficult, the GMM estimator allows to accurately identify
 the parameters. Moreover, our results indicate that the accuracy of the estimation
 increases when moments of order higher than two are included. 
  A general strategy could be to start with $k=q$ cost functions (equal to the number
of unknown parameters) and increase the maximal order
until  tests for over-identifying restrictions (e.g. the Hansen test~\cite{hansen1982large}) suggest that higher orders do not lead to an
improvement. In this way, cost functions that do not improve the quality of the estimation,
 such as the fourth order cost functions for the results in Fig.~\ref{fig:EXS}, can be identified.\\
 
We also found that an accurate approximation of the moments is crucial for 
the performance of the GMM estimator. Thus, hybrid approaches such as
the method of conditional moments~\cite{Hasenauer2013} 
or sophisticated closure schemes (e.g.~\cite{bogomolov2015adaptive}) should be preferred.
If all propensities in the   network are linear, the moment equations are exact and 
model misspecification is not an issue. However, for most networks the moments can only be
approximated, since the propensities are nonlinear, and hence the model is potentially misspecified. 
Again, statistical tests can be used to detect model misspecification~\cite{hall2004generalized}
and equations for  higher order moments may be added to the (conditional) moment equations to
 improve the approximation of the lower order moments. 
\\
Finally, we note that the GMM framework can also be
applied when the observed molecular counts are sub-
ject to measurement errors. It is straight forward to
extend the GMM framework to the case of samples
$Y_\ell + \varepsilon$ where the error term $\varepsilon$ is independent and nor-
mally distributed with mean zero.

\section{Conclusion}
Parameter inference for stochastic models of cellular processes 
demands huge computational resources. The proposed  
approach  based on the generalized method 
of moments is based on an adjustment of the 
statistical moments of the model and therefore does
not require the computation of likelihoods. 
This makes the approach appealing for complex networks where stochastic effects play an important role, since the integration
of the moment equations is typically fast compared to other 
computations such as the computation of likelihoods.
The method does not make any assumptions about the distribution
of the process (e.g. Gaussian) and complements the existing
moment-based analysis approaches in a natural way.
 
Here, we used a multistart gradient-based minimization scheme, but
the approach can be combined with any global optimization method.
We found that the weights of the cost functions computed by the GMM estimator yield 
 clearly more accurate results than trivial (identical) weights.
 In particular, the variance of the estimator decreases when moments
 of higher order are considered. 
 We focused on the estimation of reaction rate constants 
 and, as future work, we plan to investigate how well Hill coefficients
 and initial conditions are estimated. 
 
 An important advantage of the proposed method is that   in the economics literature  the  properties
  of  GMM estimators have been investigated in detail over decades and several 
  variants  and related statistical tests
  are available. 
 We will also check how accurate approximations for the 
 variance of the GMM estimator are~\cite{hall2004generalized}.
 Since we found that when moments of order higher than three are
 included, the results become slightly worse, we will in addition 
 explore the usefulness of statistical tests for over-identifying moment conditions. 
 In this way, we can ensure that only  
 moments conditions are included that improve the estimation.

\subsection*{Competing interests}
   The authors declare that they have no competing interests.

\subsection*{Acknowledgements}
This research has been funded by the German Research Council (DFG)
 as part of the Cluster of Excellence on Multimodal Computing and Interaction
 at Saarland University.
 
 \subsection*{Author's contributions}
    VW conceived the idea, initiated the work and  drafted the manuscript. AL implemented the approach, designed and performed the experiments and analyzed and visualized the results. Both authors read and approved the final manuscript.
\newpage 
 \bibliographystyle{unsrt} 

\begin{thebibliography}{10}

\bibitem{mcquarrie}
D.~A. McQuarrie.
\newblock Stochastic approach to chemical kinetics.
\newblock {\em Journal of Applied Probability}, 4:413--478, 1967.

\bibitem{MLEJournal}
Aleksandr Andreychenko, Linar Mikeev, David Spieler, and Verena Wolf.
\newblock Approximate maximum likelihood estimation for stochastic chemical
  kinetics.
\newblock {\em EURASIP J. Bioinform. Syst. Biol.}, 9, 2012.

\bibitem{milner2013moment}
Peter Milner, Colin~S Gillespie, and Darren~J Wilkinson.
\newblock Moment closure based parameter inference of stochastic kinetic
  models.
\newblock {\em Stat. Comput.}, 23(2):287--295, 2013.

\bibitem{singh2006lognormal}
Abhyudai Singh and Joao~Pedro Hespanha.
\newblock {Lognormal moment closures for biochemical reactions}.
\newblock In {\em Decision and Control, 2006 45th IEEE Conference on}, pages
  2063--2068. IEEE, 2006.

\bibitem{StumpfJournal}
A~Ale, P~Kirk, and M~P~H Stumpf.
\newblock {A general moment expansion method for stochastic kinetic models}.
\newblock {\em J. Chem. Phys.}, 138(17):174101, 2013.

\bibitem{Engblom}
S~Engblom.
\newblock {Computing the moments of high dimensional solutions of the master
  equation}.
\newblock {\em Appl. Math. Comput.}, 180(2):498--515, 2006.

\bibitem{Andreychenko2015}
Alexander Andreychenko, Linar Mikeev, and Verena Wolf.
\newblock {Model Reconstruction for Moment-Based Stochastic Chemical Kinetics}.
\newblock {\em ACM TOMACS}, 25(2):1--19, 2015.

\bibitem{CMSB10}
T.~Henzinger, M.~Mateescu, L.~Mikeev, and V.~Wolf.
\newblock Hybrid numerical solution of the chemical master equation.
\newblock In {\em Proc. of CMSB'10}, ~\!\!, 2010. ACM DL.

\bibitem{menz2012hybrid}
Stephan Menz, Juan~C Latorre, Christof Schutte, and Wilhelm Huisinga.
\newblock Hybrid stochastic--deterministic solution of the chemical master
  equation.
\newblock {\em SIAM MMS}, 10(4):1232--1262, 2012.

\bibitem{Hellander2007100}
Andreas Hellander and Per Lötstedt.
\newblock Hybrid method for the chemical master equation.
\newblock {\em J. Comput. Phys.}, 227(1):100 -- 122, 2007.

\bibitem{jahnke2011reduced}
Tobias Jahnke.
\newblock On reduced models for the chemical master equation.
\newblock {\em SIAM MMS}, 9(4):1646--1676, 2011.

\bibitem{Hasenauer2013}
J.~Hasenauer, V.~Wolf, A.~Kazeroonian, and F.~J. Theis.
\newblock Method of conditional moments ({MCM}) for the chemical master
  equation.
\newblock {\em J. Math. Biol.}, 69(3):687--735, 2013.

\bibitem{hansen1982large}
Lars~Peter Hansen.
\newblock Large sample properties of generalized method of moments estimators.
\newblock {\em Econometrica}, pages 1029--1054, 1982.

\bibitem{hall2005generalized}
Alastair~R Hall et~al.
\newblock {\em Generalized method of moments}.
\newblock Oxford University Press Oxford, ~\!\!, 2005.

\bibitem{gillespie77}
D.~T. Gillespie.
\newblock Exact stochastic simulation of coupled chemical reactions.
\newblock {\em J. Phys. Chem.}, 81(25):2340--2361, 1977.

\bibitem{sliding}
T.~Henzinger, M.~Mateescu, and V.~Wolf.
\newblock Sliding window abstraction for infinite {M}arkov chains.
\newblock In {\em Proc. CAV}, volume 5643 of {\em LNCS}, ~\!\!, 2009. Springer.

\bibitem{Munsky06}
B.~Munsky and M.~Khammash.
\newblock The finite state projection algorithm for the solution of the
  chemical master equation.
\newblock {\em J. Chem. Phys.}, 124:044144, 2006.

\bibitem{FAUIET}
M.~Mateescu, V.~Wolf, F.~Didier, and T.A. Henzinger.
\newblock Fast adaptive uniformisation of the chemical master equation.
\newblock {\em IET Syst. Biol.}, 4(6):441--452, 2010.

\bibitem{Inexact}
R.~Sidje, K.~Burrage, and S.~MacNamara.
\newblock Inexact uniformization method for computing transient distributions
  of {M}arkov chains.
\newblock {\em SIAM J. Sci. Comput.}, 29(6):2562--2580, 2007.

\bibitem{van1992stochastic}
Nicolaas~Godfried Van~Kampen.
\newblock {\em Stochastic processes in physics and chemistry}, volume~1.
\newblock Elsevier, ~\!\!, 1992.

\bibitem{thomas2015}
Philipp Thomas and Ramon Grima.
\newblock Approximate probability distributions of the master equation.
\newblock {\em Phys. Rev. E}, 92(1):012120, 2015.

\bibitem{Timmer}
S.~Reinker, R.M. Altman, and J.~Timmer.
\newblock Parameter estimation in stochastic biochemical reactions.
\newblock {\em IEEE Proc. Syst. Biol}, 153:168--178, 2006.

\bibitem{schnoerr2015}
David Schnoerr, Guido Sanguinetti, and Ramon Grima.
\newblock Comparison of different moment-closure approximations for stochastic
  chemical kinetics.
\newblock {\em J. Chem. Phys.}, 143(18):185101, 2015.

\bibitem{bogomolov2015adaptive}
Sergiy Bogomolov, Thomas~A Henzinger, Andreas Podelski, Jakob Ruess, and
  Christian Schilling.
\newblock Adaptive moment closure for parameter inference of biochemical
  reaction networks.
\newblock In {\em Proc. of CMSB'15}, pages 77--89. Springer International
  Publishing.

\bibitem{grima2012study}
Ramon Grima.
\newblock A study of the accuracy of moment-closure approximations for
  stochastic chemical kinetics.
\newblock {\em J. Chem. Phys.}, 136(15):154105, 2012.

\bibitem{Soltani2015}
M.~Soltani, C.~A. Vargas-Garcia, and A.~Singh.
\newblock Conditional moment closure schemes for studying stochastic dynamics
  of genetic circuits.
\newblock {\em IEEE Transactions on Biomedical Circuits and Systems},
  9(4):518--526, Aug 2015.

\bibitem{kazeroonian2016cerena}
Atefeh Kazeroonian, Fabian Fr{\"o}hlich, Andreas Raue, Fabian~J Theis, and Jan
  Hasenauer.
\newblock {CERENA}: Chemical reaction network analyzer: A toolbox for the
  simulation and analysis of stochastic chemical kinetics.
\newblock {\em PloS one}, 11(1):e0146732, 2016.

\bibitem{gillespie76}
D.~T. Gillespie.
\newblock A general method for numerically simulating the time evolution of
  coupled chemical reactions.
\newblock {\em J. Comput. Phys.}, 22:403--434, 1976.

\bibitem{ruess2015minimal}
Jakob Ruess.
\newblock Minimal moment equations for stochastic models of biochemical
  reaction networks with partially finite state space.
\newblock {\em J. Chem. Phys.}, 143(24):244103, 2015.

\bibitem{HanleyLomasMittarMainoPark2013}
Mary~Beth Hanley, Woodrow Lomas, Dev Mittar, Vernon Maino, and Emily Park.
\newblock Detection of low abundance {RNA} molecules in individual cells by
  flow cytometry.
\newblock {\em PLoS ONE}, 8(2):1--8, 02 2013.

\bibitem{Hasenauer2011}
Jan Hasenauer, Steffen Waldherr, Malgorzata Doszczak, Nicole Radde, Peter
  Scheurich, and Frank Allg{\"o}wer.
\newblock Identification of models of heterogeneous cell populations from
  population snapshot data.
\newblock {\em BMC Bioinformatics}, 12(1):1--15, 2011.

\bibitem{Munsky201512}
Brian Munsky, Zachary Fox, and Gregor Neuert.
\newblock Integrating single-molecule experiments and discrete stochastic
  models to understand heterogeneous gene transcription dynamics.
\newblock {\em Methods}, 85:12 -- 21, 2015.
\newblock Inferring Gene Regulatory Interactions from Quantitative
  High-Throughput Measurements.

\bibitem{hall2004generalized}
A.R. Hall.
\newblock {\em Generalized Method of Moments}.
\newblock Advanced Texts in Econometrics. OUP Oxford, ~\!\!, 2004.

\bibitem{loinger-lipshtat-balaban-biham07}
A.~Loinger, A.~Lipshtat, N.~Q. Balaban, and O.~Biham.
\newblock Stochastic simulations of genetic switch systems.
\newblock {\em Phys. Rev. E}, 75:021904, 2007.

\bibitem{hscc11}
Maksim Lapin, Linar Mikeev, and Verena Wolf.
\newblock {SHAVE} -- {S}tochastic hybrid analysis of {M}arkov population
  models.
\newblock In {\em Proc. of HSCC'11}, ACM International Conference Proceeding
  Series, 2011.

\bibitem{ugray2007scatter}
Zsolt Ugray, Leon Lasdon, John Plummer, Fred Glover, James Kelly, and Rafael
  Mart{\'\i}.
\newblock Scatter search and local nlp solvers: A multistart framework for
  global optimization.
\newblock {\em INFORMS JOC}, 19(3):328--340, 2007.

\bibitem{hansen1996finite}
Lars~Peter Hansen, John Heaton, and Amir Yaron.
\newblock Finite-sample properties of some alternative {GMM} estimators.
\newblock {\em J. Bus. Econ. Stat.}, 14(3):262--280, 1996.

\bibitem{Wilkinson2}
R.~Boys, D.~Wilkinson, and T.~Kirkwood.
\newblock Bayesian inference for a discretely observed stochastic kinetic
  model.
\newblock {\em Stat. Comp.}, 18:125--135, 2008.

\bibitem{wilkinson2011stochastic}
Darren~J Wilkinson.
\newblock {\em Stochastic modelling for systems biology}.
\newblock CRC press, ~\!\!, 2011.

\bibitem{golightly2011bayesian}
Andrew Golightly and Darren~J Wilkinson.
\newblock Bayesian parameter inference for stochastic biochemical network
  models using particle markov chain monte carlo.
\newblock {\em Interface focus}, page rsfs20110047, 2011.

\bibitem{daigle2012accelerated}
Bernie~J Daigle, Min~K Roh, Linda~R Petzold, and Jarad Niemi.
\newblock Accelerated maximum likelihood parameter estimation for stochastic
  biochemical systems.
\newblock {\em BMC Bioinformatics}, 13(1):68, 2012.

\bibitem{drovandi2011estimation}
Christopher~C Drovandi and Anthony~N Pettitt.
\newblock Estimation of parameters for macroparasite population evolution using
  approximate bayesian computation.
\newblock {\em Biometrics}, 67(1):225--233, 2011.

\bibitem{fearnhead2012constructing}
Paul Fearnhead and Dennis Prangle.
\newblock Constructing summary statistics for approximate bayesian computation:
  semi-automatic approximate bayesian computation.
\newblock {\em J. R. Stat. Soc. Series B Stat. Methodol.}, 74(3):419--474,
  2012.

\bibitem{Stumpf}
T.~Toni, D.~Welch, N.~Strelkowa, A.~Ipsen, and M.~Stumpf.
\newblock {Approximate Bayesian computation scheme for parameter inference and
  model selection in dynamical systems}.
\newblock {\em J. R. Soc. Interface}, 6(31):187--202, 2009.

\bibitem{Tian2007}
T.~Tian, S.~Xu, J.~Gao, and K.~Burrage.
\newblock Simulated maximum likelihood method for estimating kinetic rates in
  gene expression.
\newblock {\em Bioinformatics}, 23:84--91, 2007.

\bibitem{Uz2010478}
B.~Uz, E.~Arslan, and I.~Laurenzi.
\newblock Maximum likelihood estimation of the kinetics of receptor-mediated
  adhesion.
\newblock {\em J. Theor. Biol.}, 262(3):478 -- 487, 2010.

\bibitem{Higgins}
J.~J. Higgins.
\newblock Bayesian inference and the optimality of maximum likelihood
  estimation.
\newblock {\em Int. Stat. Rev.}, 45(1):9--11, 1977.

\bibitem{zechner2012moment}
Christoph Zechner, Jakob Ruess, Peter Krenn, Serge Pelet, Matthias Peter, John
  Lygeros, and Heinz Koeppl.
\newblock Moment-based inference predicts bimodality in transient gene
  expression.
\newblock {\em PNAS}, 109(21):8340--8345, 2012.

\bibitem{ruess2015moment}
Jakob Ruess and John Lygeros.
\newblock Moment-based methods for parameter inference and experiment design
  for stochastic biochemical reaction networks.
\newblock {\em ACM TOMACS}, 25(2):8, 2015.

\bibitem{kugler2012moment}
Philipp K{\"u}gler.
\newblock Moment fitting for parameter inference in repeatedly and partially
  observed stochastic biological models.
\newblock {\em PloS one}, 7(8):e43001, 2012.

\bibitem{Hasenauer2016}
Fabian Fr\"ohlich, Philipp Thomas, Atefeh Kazeroonian, Fabian~J. Theis, Ramon
  Grima, and Jan Hasenauer.
\newblock Inference for stochastic chemical kinetics using moment equations and
  system size expansion.
\newblock {\em PLoS Comput Biol}, 12(7):1--28, 07 2016.

\bibitem{Komorowski2009}
Micha{\l} Komorowski, B{\"a}rbel Finkenst{\"a}dt, Claire~V. Harper, and
  David~A. Rand.
\newblock Bayesian inference of biochemical kinetic parameters using the linear
  noise approximation.
\newblock {\em BMC Bioinformatics}, 10(1):1--10, 2009.

\bibitem{zimmer2015deterministic}
Christoph Zimmer and Sven Sahle.
\newblock Deterministic inference for stochastic systems using multiple
  shooting and a linear noise approximation for the transition probabilities.
\newblock {\em Systems Biology, IET}, 9(5):181--192.

\bibitem{bergmann2016piecewise}
Frank~T Bergmann, Sven Sahle, and Christoph Zimmer.
\newblock Piecewise parameter estimation for stochastic models in copasi.
\newblock {\em Bioinformatics}, page btv759, 2016.

\end{thebibliography}

\newpage 
\begin{appendix}
\section{Influence of multiple Time Points}\label{app}
\begin{figure*}[h]
    \includegraphics[width=0.99\textwidth]{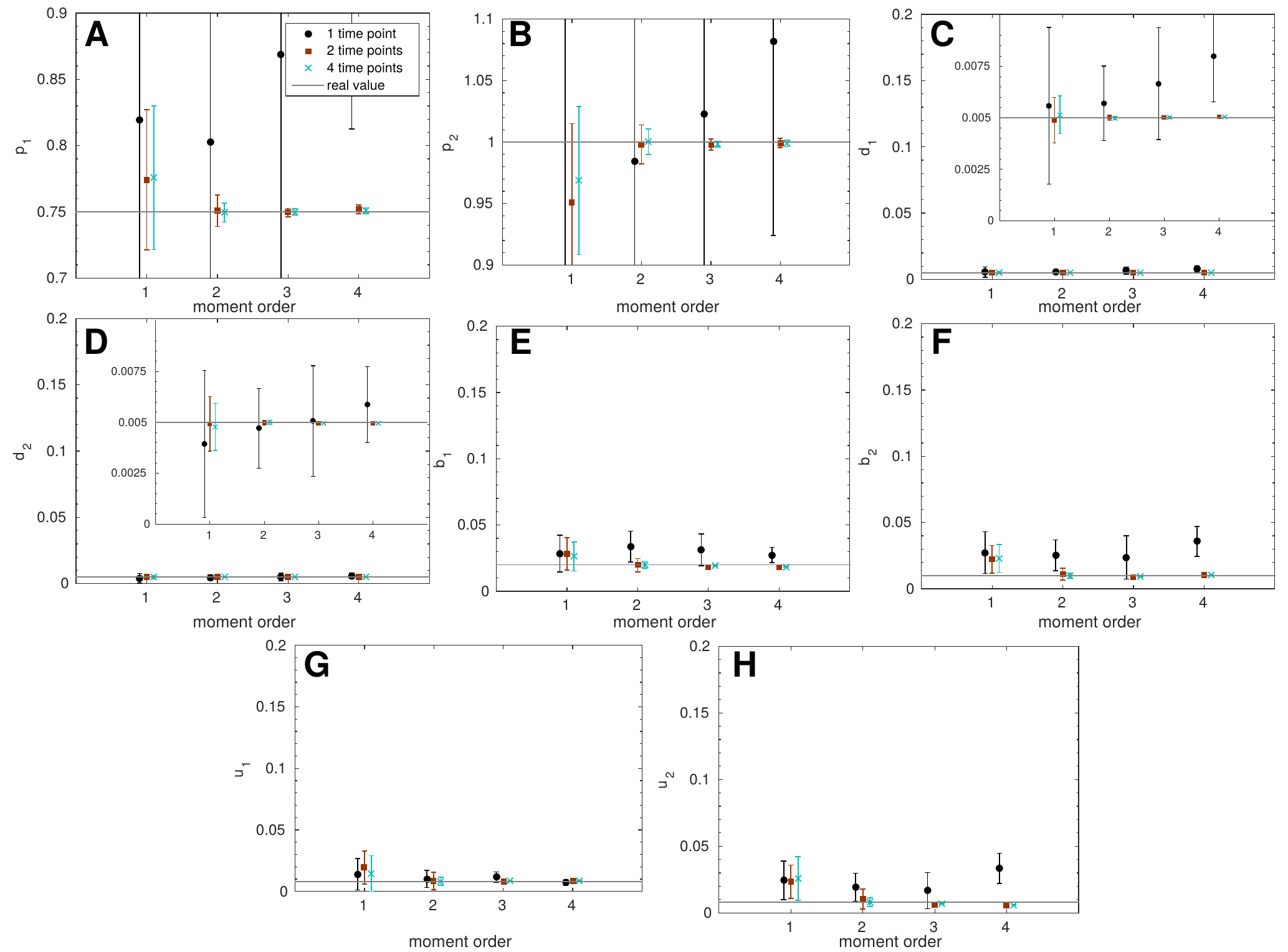}
    \caption{Exclusive switch model: Comparison of estimations with the demean procedure for single time point data and combined data for samples of two and four independent time points.}\label{fig:diffTimes}
\end{figure*}

For certain pairs of chemical reactions the identifying condition $$
 E[{\bf f}(Y,\theta)]={\bf 0} \mbox{ if and only if } \theta=\theta_0.
 $$ is violated when only regarding snapshot data from a single time point. For example in the very simple reaction system
\begin{align*}
\emptyset &\rightarrow A~~\text{rate~}\lambda, \\
A&\rightarrow \emptyset~~~\text{rate~}\mu
\end{align*} 
every combination of $\lambda$ and $\mu$ with $\frac{\lambda}{\mu}=const$ would lead to the same snapshot data for species $A$ at a certain time point.\\
In order to resolve this problem more information, i.e. snapshot data at several time points (of independent samples to avoid correlation), is needed or one of the parameters has to be fixed. In section~\ref{sec:SVH} this problem already occured for the exclusive switch: The corresponding rates are production $p_i$ and degradation $d_i$ as well as binding $b_i$ and unbinding $u_i$. By fixing the degredation rates $d_i$ the estimation of the production rates becomes quite well, whereas $b_i$ und $u_i$ can not be estimated due to the identifying problem.\\
For the following estimations the demean procedure was used. The 2-Step method showed a similar behavior.
With no fixed parameters and only a single time point $t=200$ nothing can be reliably estimated as indicated in Fig.~\ref{fig:diffTimes}. The estimated values are often far away from the real ones and the variance is also quite high in all cases.\\
The consideration of a second time point ($t=100,200$) resolves the issue in case of sufficient moment conditions, i.e. order 2 or higher. Four time points ($t=50, 100, 150, 200$) do not further improve the estimation but due to the higher total number of samples ($500{,}000$ per time point) the variance is decreased.
\end{appendix}
\end{document}